# A Computational Framework Integrating Physics-based Model and Equivalent Circuit Network Model to Simulate Li-ion batteries

*Han Yuan[1]*, Shen Li[1], Tao Zhu[1], Simon O'Kane[1], Carlos Garcia[1], Gregory Offer[1,2], Monica Marinescu[1,2]*

*1 Department of Mechanical Engineering Imperial College London, London, SW7 2AZ, UK*

*2 The Faraday Institution, Harwell Science and Innovation Campus, Didcot OX11 0RA, United Kingdom*

*Corresponding author email: han.yuan16@imperial.ac.uk*

## Abstract

Computational models of Li-ion batteries are of great importance in the cell/pack design, operation, and control. Battery models generally fall into two categories: physics-based models and ECM models. Physics-based Doyle-Fuller-Newman (DFN) models can accurately simulate the battery internal electrochemical processes, but to properly account for thermal effects requires a strong coupling between a DFN model and a 3D thermal model, which is computationally unaffordable. Distributed Equivalent Circuit Network (ECN) models can perform simulations with high speed and reasonable accuracy. However, these models rely heavily on the characterisation experiments for ECN parameter identification, which is resource-intensive and can lead to inaccurate parametrisation outcomes due to internal thermal inhomogeneity. To harness the strengths of both models, we propose a computational framework to integrate electrochemical DFN model and 3D distributed ECN model together. The workflow of the framework consists of three steps: the characterisation data acquisition from DFN modelling, ECM parametrisation, and 3D distributed ECN modelling. Using this framework, we simulate constant current discharge experiments of Kokam 7.5 Ah pouch cell (Model SLPB75106100) and compare the simulations with the commonly-used lumped DFN-thermal model. The computational model outperforms the lumped DFN model at low-temperature and/or high C-rate scenarios significantly. The largest predicting error of the framework at 3 C-rate &$T_{am}$ = 25°C and at 1 C-rate &$T_{am}$ = 0 °C is approximately 1/3 of that for DFN model. At 3 C-rate &$T_{am}$ = 5°C, the difference between these two can rise to 377 mV. Further analysis reveals that it is of great importance to properly account for the thermal effects during cell cycling and, the commonly-used lumped DFN-thermal model is not suitable to simulate the performance of Li-ion batteries, when the internal heat generation and Biot number are large. By integrating DFN and 3D-distributed ECN together, the computational framework can simulate the complicated interplay between electrochemistry, thermal process, and electricity within a cell fast and accurately. We anticipate this computational framework to be a valuable toolset to assist researchers and engineers in the design and control of Li-ion batteries.

**Keywords:** Lithium-ion battery; Electrochemical-Electro-thermal Coupling; Computational study; DFN modelling, ECM modelling

## 1 Introduction

Since commercialised by Sony Corporation in 1991 [1], Li-ion batteries have been widely used in electronic devices, transportation, and energy storage systems. Due to the concern over carbon emissions and climate change, governments around the world launched initiatives on electric vehicles and renewable energy, the demand of Li-ion batteries has been increasing drastically. It is projected that by 2030, the global demand of lithium-ion batteries will reach 1500 GWh, 10-time larger than that in 2020 [2]. The ongoing research in the community seeks to further improve the performance, affordability, safety, and lifetime of Li-ion batteries. Modelling is a key tool to achieve these improvements, as it is of great importance in the battery design, operation, and control. Battery models generally fall into two categories: physics-based models and ECM models.

The physics-based electrochemical battery models were initially developed in the 1960s and 70s [3–5] by John Newman, who developed the theories of *concentrated solution* and *porous electrode*. Built on these fundamentals, in the middle of 90s, Doyle, Fuller and Newman presented a model to simulate charge and discharge of Li-ion cells [6,7], known as DFN (Doyle-Fuller-Newman) model. The DFN model assumes at each point of the electrode there is a spherical particle which is representative of the active material. Thus it considers one spatial dimension for the electrode thickness and another spatial dimension for the radius of each particle. It is therefore also referred to as P2D (pseudo-two-dimensional) model. DFN model charaterises the consequence of complicated interrelated phenomena during the cycling of a Li-ion cell. Appling several coupled partial differential equations along with specified boundary conditions to mathematically describe those phenomena, DFN model demonstrates a high accuracy in representing the battery internal electrochemical processes, well predicting the potential and current distributions in both the porous electrodes and the electrolyte, the lithium ion concentration within the electrolyte and the distribution of intercalated lithium within the electrode particles [8–11]. It has been used as a efficient design tool in battery industry to faciliate the development of new eletrode and cell by minimising the need of expensive iterations [11].

However, the DFN model has its limitations. Firstly, it is computationally expensive in nature. Solving a system of partial differential equations that underpins DFN model requires considerable computing time and resource. In practice, the DFN modelling is typically based on 1D assumptions [8–11], where dimensionality is sacrificed for lower computational cost. Yet, for large-format cells or cell packs, where the dimension effect becomes significant, those assumptions does not hold true. Moreover, to fully couple the DFN with a 3D thermal model is computationally unaffordable in practice. Due to internal electrochemical processes (such as entropy change in a reaction, overpotentials, the resistance during ion transportation, and resistance of current flow) , the cycling of a cell can generate/absorb a significant amount of heat [12]. On the other hand, physicochemical properties are strongly dependent on temperature (in particular diffusivities, reaction rates and the electrolyte conductivity). For example, the diffusivity of Li-ion increases with temperature due to the increased kinetic energy, and interface reaction rate also accelerates at high temperatures as described in the Bulter-Volmer equation. The increase or decline of cell temperature leads to the change of electrochemical properties, which in turn affect the electrochemical processes. What makes the problem even more complicated is the chemical and thermal inhomogeneity within the cell. Temperature [13], lithium concentration [14], and current density [15] can vary significantly across the geometrical structure of a cell [13–15]. Therefore, to accurately simulate the interplay between the electrochemical and thermal processes with a DFN model requires a strong coupling with a 3D thermal model, which is considered to be too costly and time-consuming to apply in engineering practice [11].

To mitigate the intense computational demands, Equivalent Circuit models (ECMs) were proposed to be used for the simulations of Li-ion batteries approximately two decades ago [16,17]. As empirical models, ECMs do not reply on the detailed first-principle-based physics. Instead, they assume that the battery can be represented by an electrical circuit, typically comprised of resistors and capacitors, and then fit the parameters of the circuit components to experimental data [17,18]. Due to their empirical nature, ECMs can perform faster calculation than physics-based models, and therefore are primarily used in real-time control and management of battery system, where the simulation speed is essential.

The original ECM doesn't incorporate thermal processes. To account for the thermal effects and its coupling with other mechanisms, our research group proposed to couple ECM model with thermal model and to split a whole cell into different ECM units to form a computational network. This approach led to the successful construction of multi-dimensional, electro-thermally coupled, distributed Equivalent Circuit Network (ECN) models for cells with 3 different form factors, namely pouch [19,20], cylindrical [13,21], and prismatic [22]. Integrated the abovementioned research together, we recently developed PyECN (Python Equivalent Circuit Network), a tool crafted using the open-source Python language [23]. PyECN allows for the discretization of a cell's geometrical structure into a number of computing units, and each computing units contains both electrical and thermal sub-units. The electrical and thermal sub-units are strongly coupled to simulate the complicated electro-thermal behaviour of batteries.

Due to the progress in distributed ECN models, ECMs has demonstrated the capability to conduct fast computing, while maintaining a reasonable accuracy. However, ECMs are not without their limitations. In general, ECMs are models fitted from a series of experimental data sets generated from characterisation tests at different operating conditions [17,22,24,25]. As the parameters of ECMs vary with operating conditions [26,27], it is necessary for the characterisation tests to encompass a broad spectrum of operating variables such as current, temperature, and cycle numbers [24], which is highly resource-intensive and time-consuming. Moreover, the thermal processes during characterisation tests influence the parametrisation outcomes [28]. The characterisation tests are typically conducted using commercialised cells. In the thermal chamber, the constant temperature condition is maintained by measuring cell surface temperature and/or chamber temperature [13,22]. However, heat generation and temperature inhomogeneity arise during discharge and charge of a cell, which can be very significant when the cell is large-format [27] or during long-term discharge/charge [21]. Often, the cell's average temperature is higher than the controlled value, with temperature difference within the cell reaching up to 10 ºC [21]. Consequently, the parameters of ECM derived from a specified chamber temperature may not accurately reflect the cell's electrochemistry at that temperature, given that different layers and components of a cell can have markedly different temperatures.

It is seen that both physics-based and ECMs models their own advantages and limitations. To combine the best aspects of these two types of models, this work presents a computational framework that integrates DFN model and PyECN together, aiming to deliver fast and accurate simulations, whereas minimise the experimental resources. The framework is validated against experiments of constant current discharge at various operating conditions (i.e different C-rates and ambient temperatures). A comparative analysis is performed between this computational framework and the commonly-used lumped DFN-thermal model. This comparison underscores the advantages of our framework and highlights the importance of properly accounting for thermal effects in Li-ion battery simulations.

## 2 Methodology

Fig. 1 shows the computational framework and corresponding workflow. The framework consists of three components： DFN (Doyle-Fuller-Newman) modelling, ECM (Equivalent Circuit Model) parametrisation, and 3D distributed ECN (Equivalent Circuit Network) modelling. In our framework, a parametrised DFN model is implemented as a virtual experimental rig at coin-cell level to produce characterisation data at various conditions (C-rates, temperatures, SoCs), which is the first step of the workflow. The second step is ECM parametrisation. The characterisation data generated from DFN model are fed into a parametrisation code to generate ECM parameters through inverse-modelling optimisation algorithm. Finally, the ECM parameters along with the thermal properties and geometries of a cell will be given to PyECN as input to simulate the performance of a cell. PyECN is 3D distributed electro-thermal coupled ECN (Equivalent Circuit Network) model created in our group [13,23,25], and its thermal sub-model can simulate the heat transfer and heat generation in a cell [13,25]. Our framework integrates DFN model and electro-thermal coupled ECN model together and it well considers the electrochemistry and thermal process as well as their coupling effects. The rest of the section will explain how the computational work works in detail using a case study and present justification to apply the computational framework as an alternative to conventional simulation approaches.

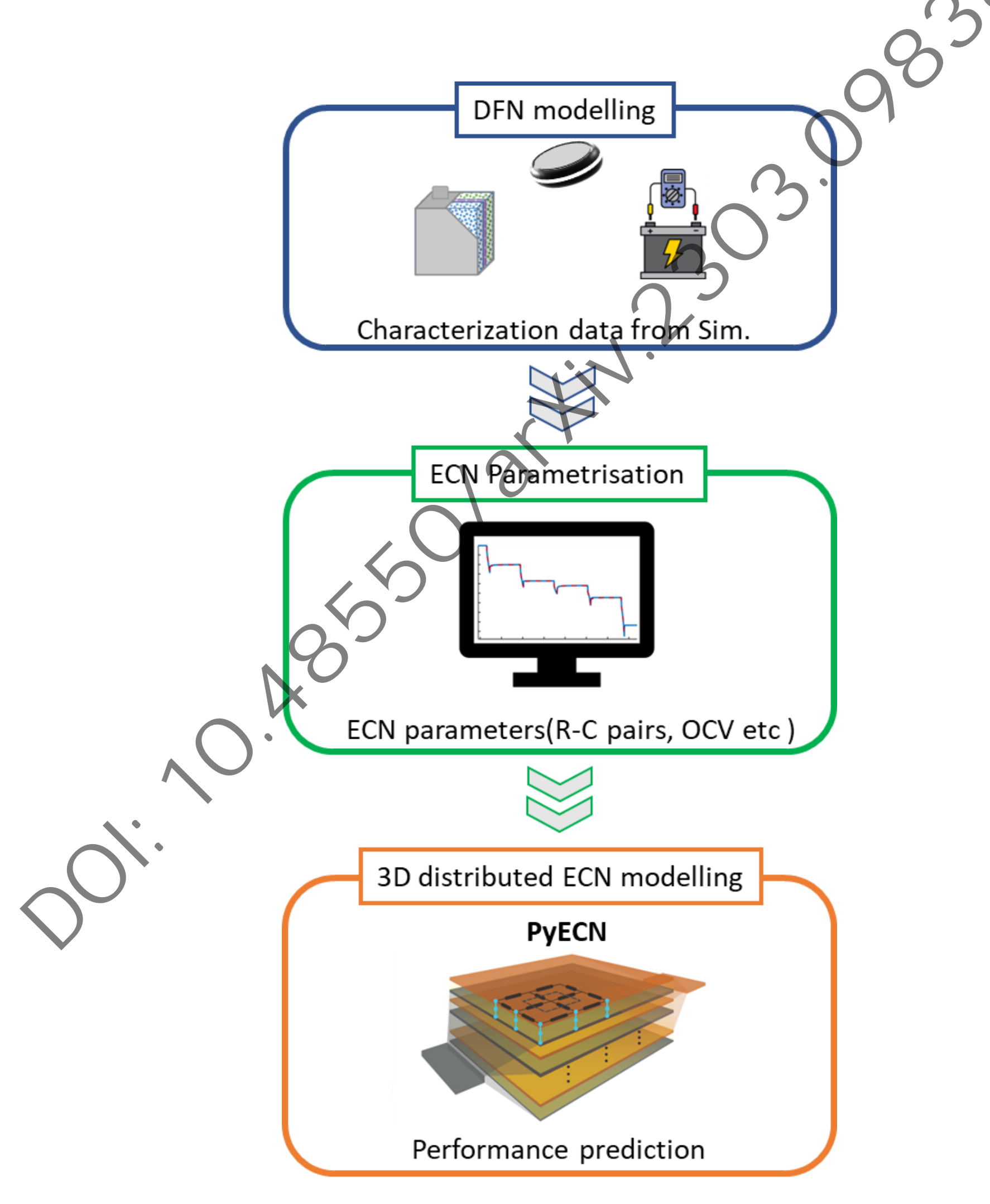

Fig. 1 The computational framework that integrates physics-based electrochemical DFN model and distributed ECN Model (PyECN)

## 2.1 DFN modelling

The first step of work flow is DFN modelling to generate characterisation data. In this paper, the implementation of DFN model was based on the open-source code PyBaMM [29] (Python Battery Mathematical Modelling) package. Our DFN modelling considers the temperature dependence of electrochemistry, while the whole cell temperature is set to be same as ambient during a single (virtual) test. The PyBaMM model here is therefore regarded as an ideal coin-cell level experimental rig. Compared to realistic experimental tests, the characterisation data sets can be produced in a much faster manner, whereas the influence of thermal inhomogeneity (discussed in the introduction) is completely eliminated. Then, the generated characterisation data at different temperatures will be applied to fit ECM parameters (Section 2.2), which will be eventually used in every electrical unit of distributed electro-thermally coupled Equivalent Circuit Network model in the third step (Section 2.3).

The cell investigated in this research is a 7.5 Ah cell produced by Kokam (Model: SLPB75106100). It has a positive electrode made of $Li(Ni_{0.4}Co_{0.6})O_2$ and negative electrode made of graphite. This cell was chosen as the parametrisation for electrochemical model has been conducted at coin-cell level [9,30] with the values reported (Table I and Table II in [9]). Based on this parametrisation, the DFN model was built straightforward for the cell.

Fig. 2 shows an example of characterisation data set generated by PyBaMM at the condition of 3 C-rate and $T_{am}$= 25°C. A single data set consists of two discharging segments: Segment 1 is the cell discharges with constant current from full until 7.5Ah (nominal discharging capacity) of charge has passed. Segment 2 is a 25-pulse discharge GITT (Galvanostatic Intermittent Titration Technique): All 25 pulses have the same stop condition: either 0.3 Ah of charge is passed or the voltage reaches 2.6 V (whichever comes first), and are separated by the same rest period of 3600 seconds. It is noteworthy that our preliminary studies found ECM parameters vary significantly with current load, aligning with the findings from other studies [31,32]. Therefore, except for temperature, different C-rates were considered. In total, data sets were generated at 14 operating condition: 2 C-rates ( 1 and 3 C-rate) × 7 operating temperatures (0 °C, 5 °C, 15 °C, 25 °C, 35 °C, 45 °C, 55 °C). Each data set corresponds to one operating condition. All the other characterisation data sets are presented in Appendix A for reference.

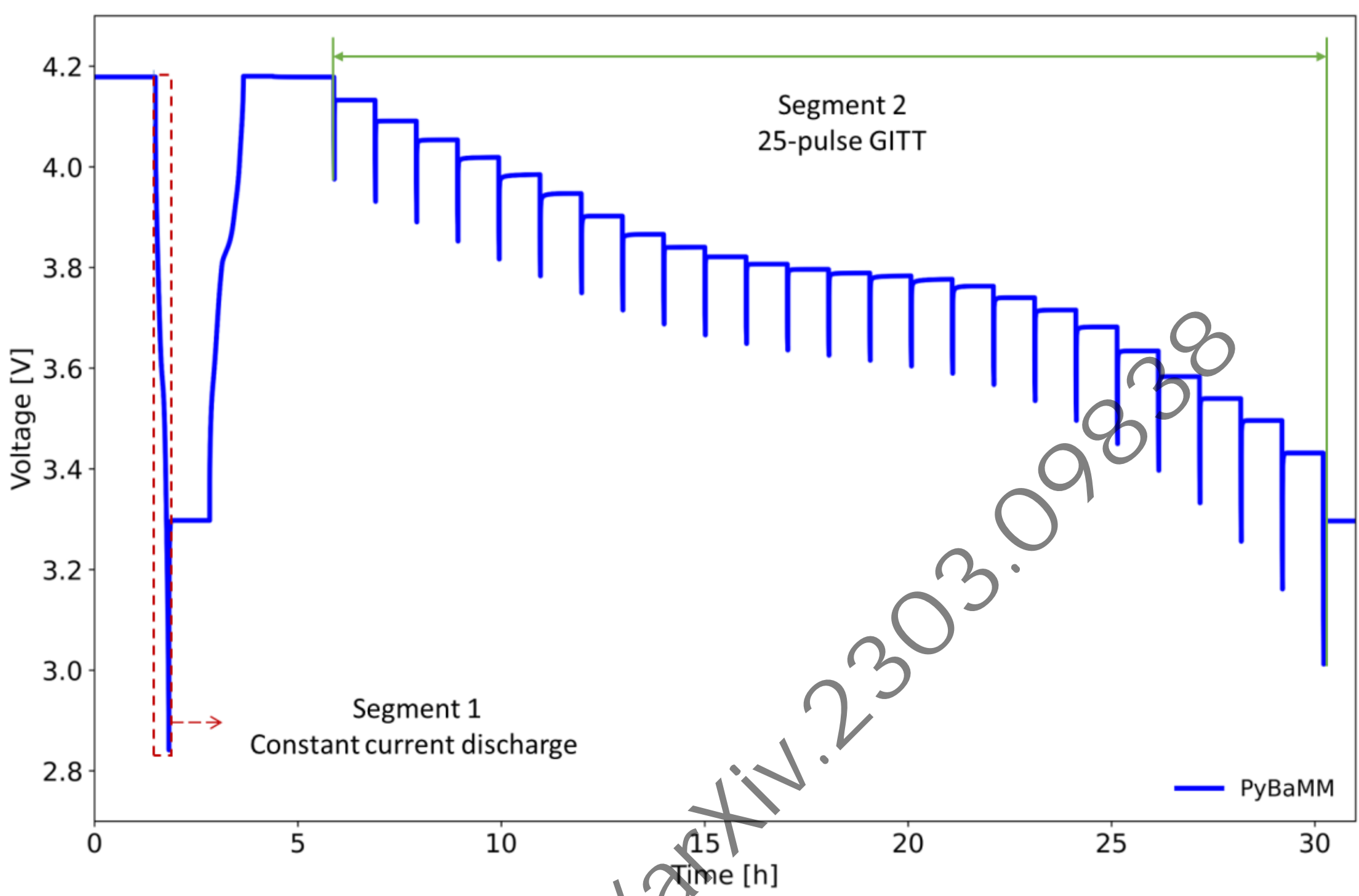

Fig. 2 Characterization data generated from PyBaMM at 3C and $T_{am}$= 25°C; A set of data consists of constant current data and GITT data

## 2.2 ECM parametrizations

The second part of the framework is the ECM parametrisation. The characterisation data obtained in Section 2.1 was utilised to produce ECM parameters, providing the link between the two models: DFN and 3D distributed ECN.

The ECM is formed by an open voltage source (OCV), a series resistance R0, and 2 R-C (resistor–capacitor in parallel) pairs: $R_i$ and $C_i$, where i = 1,2 (see Fig. 5). As standard, the OCV is the static part representing the thermodynamic potential of the cell, while $R_0$ represents the instantaneous response and the R-C pairs the dynamic components of the cell internal impedance [17]. The ECM parameters are used as inputs to the electrical unit in the 3D distributed ECN (see Fig. 5), thus allowing the prediction of battery behaviour at various operating conditions.

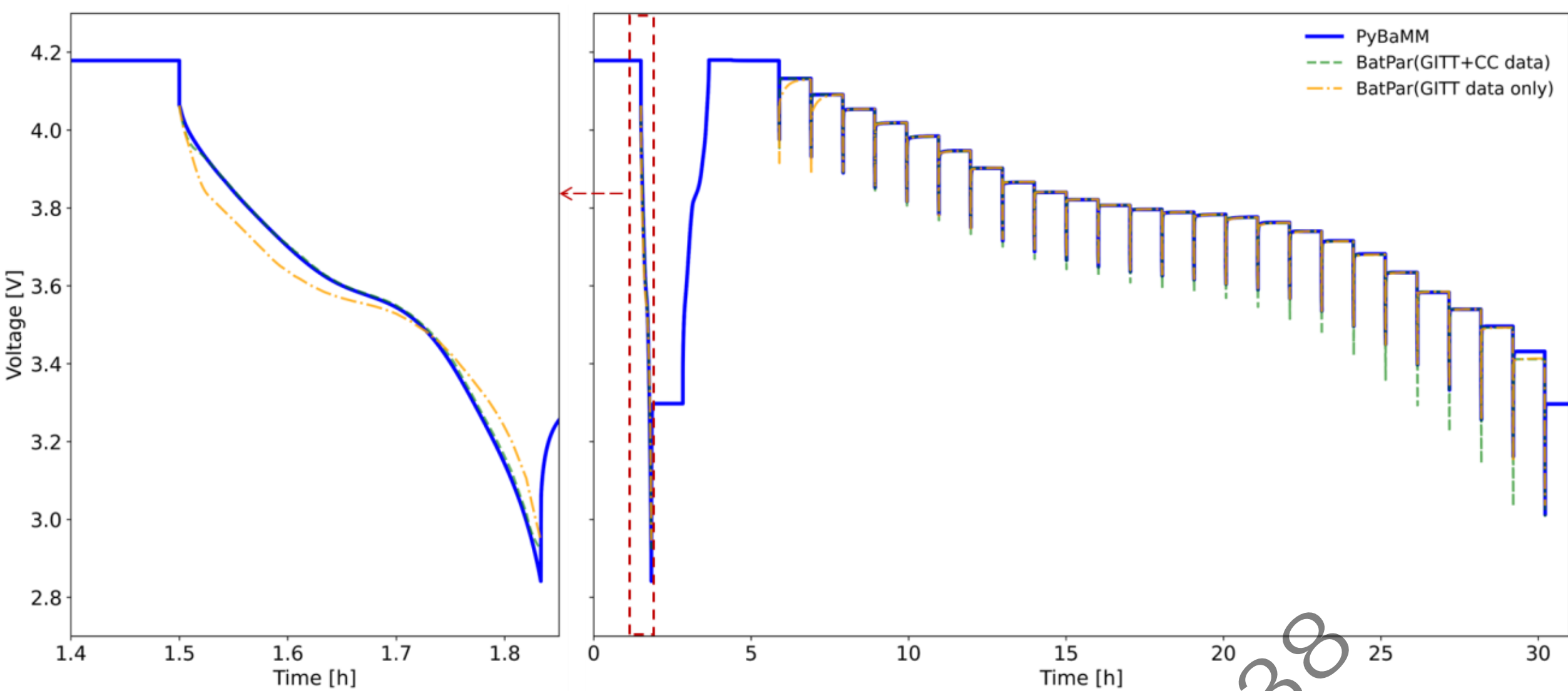

Fig. 3 PyBaMM simulation Vs the fitted curves from BatPar. The comparison shown here is at the condition of 3 C-rate and $T_{am}$= 25 °C. The solid line is PyBaMM simulation that was fed into BatPar for parametrization. The dashed green line was the fitted curve obtained by feeding constant current data and 25-pulse GITT data into BatPar simultaneously; The dashed orange line was the fitted curve obtained when only 25-short-pulse GITT data was fed.

The ECM parameterisation were conducted through a optimisation Matlab code (BatPar) created in our group[33]. All 14 characterisation data sets generated in Section 2.1 ( See Appendix A ) were parametrised. Fig. 3 demonstrates a comparison of original PyBaMM simulation and ECM simulation generated with the fitted ECM parameters at 3 C-rate and $T_{am}$=25 °C .

It is worth noting that previous studies [13,19,22] usually use GITT-type data alone for ECM parametrisation. However, our preliminary investigation found that deriving ECM parameters solely from GITT-type data can lead to overfitting on this data and results in a noticeable mismatch for constant current discharge, particularly at high C-rate scenarios (as depicted by the orange curve in Fig. 3). This manifests the influence of load profile on the ECM parametrisation [26,34]. To ensure a more robust parameter set for different situations, our characterisation data therefore included both constant current data and GITT data. With two types of data fed in, the ECM parameters can simulate both constant current discharge and GITT experiments reasonably well. It is seen in Fig. 3 that in terms of constant current discharge, the overall root-mean-square error between PyBaMM data and the fitted curve produced by BatPar is 8.43 mV (the comparison between the blue and green curve in Fig. 3 ).

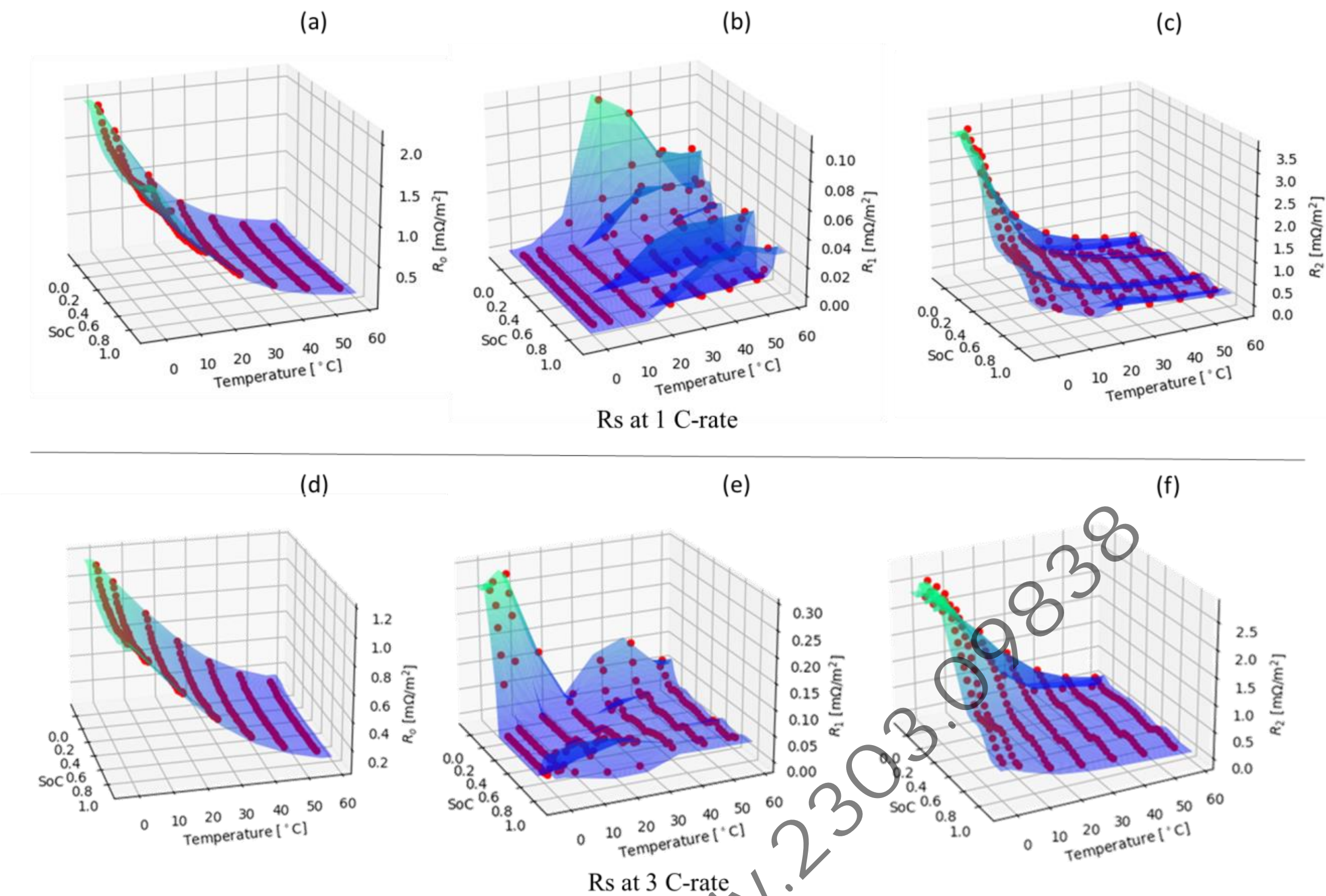

Fig. 4 $R_0$, $R_1$ and $R_2$ at different SoC and temperatures parametrised from BatPar.(a), (b), and (c) are $R_0$, $R_1$ and $R_2$ at 1 C-rate; (d), ©, and (f) are $R_0$, $R_1$ and $R_2$ at 3 C-rate. Resistances were parametrised with a SoC increment of 0.04 (SoC range from 0-1, and 26 parametrised points in total) and at 7 different temperatures (0 °C, 5 °C, 15 °C, 25 °C, 35 °C, 45 °C, 55 °C). Every parameter has 182 (26*7) parametrised points (shown as dots in the Figure). Linear interpolation was used to obtain the values between the parametrised points (shows as surfaces in the figure).

The generated ECM parameters ($R_0$, $R_1$, $R_2$, $C_1$, $C_2$) are functions of SoC, temperature, and C-rate. They were parametrised with a SoC increment of 0.04 (SoC range from 0-1, and 26 parametrised points in total) and at 7 different temperatures (0 °C, 5 °C, 15 °C, 25 °C, 35 °C, 45 °C, 55 °C). Every parameter therefore has 182 (26*7) fitting datapoints (shown as dots in the Fig. 4) Linear interpolation was used to obtain the values between the parametrised points. Fig. 4 presents the parametrised resistances (the linear interpolation is shown as the surface in the Fig. 4.). It is seen that, the cell resistance is mainly determined by $R_0$ and $R_2$ ($R_1$ is approximately one order magnitude smaller). Since $R_0$ and $R_2$ are decreasing with temperature (due to high reactivity and diffusivity at high temperatures), the overall cell resistance is higher at low temperatures.

## 2.3 3D distributed ECN model

The third part of the framework is simulation through 3D distributed electro-thermally coupled ECN model, PyECN. Fig. 5 shows the schematic representation of the 3D distributed ECN model for the pouch cell (Kokam 7.5 Ah Model: SLPB75106100) investigated in this paper. The detailed constructure and mathematical formation of PyECN can be found in [13,23]. In the model, the whole cell is divided into numerous computational units. A computational unit domain is composed of an electrode pair subdomain (i.e. anode, separator, and cathode) and two current collector subdomains. Each subdomain is represented by an electrical and a thermal ECN unit. The electrical unit(i.e ECM model) and thermal unit are fully coupled: the electrical parameters (i.e ECM parameters) are function of temperature (See Section 2.2), while heat generation from the electrical model contributes to the heat source of the thermal model. The ECN model can therefore describe the temporal and spatial evolution of various variables of interest (such as current density, temperature, SoC, heat generation) for a given load and thermal management choice. In our simulations, a distributed model of 175 electrical/thermal ECN units were used: 5 units along y axis, 5 units along z axis and 7 units along x axis. This discretization has been chosen through convergence check investigation.

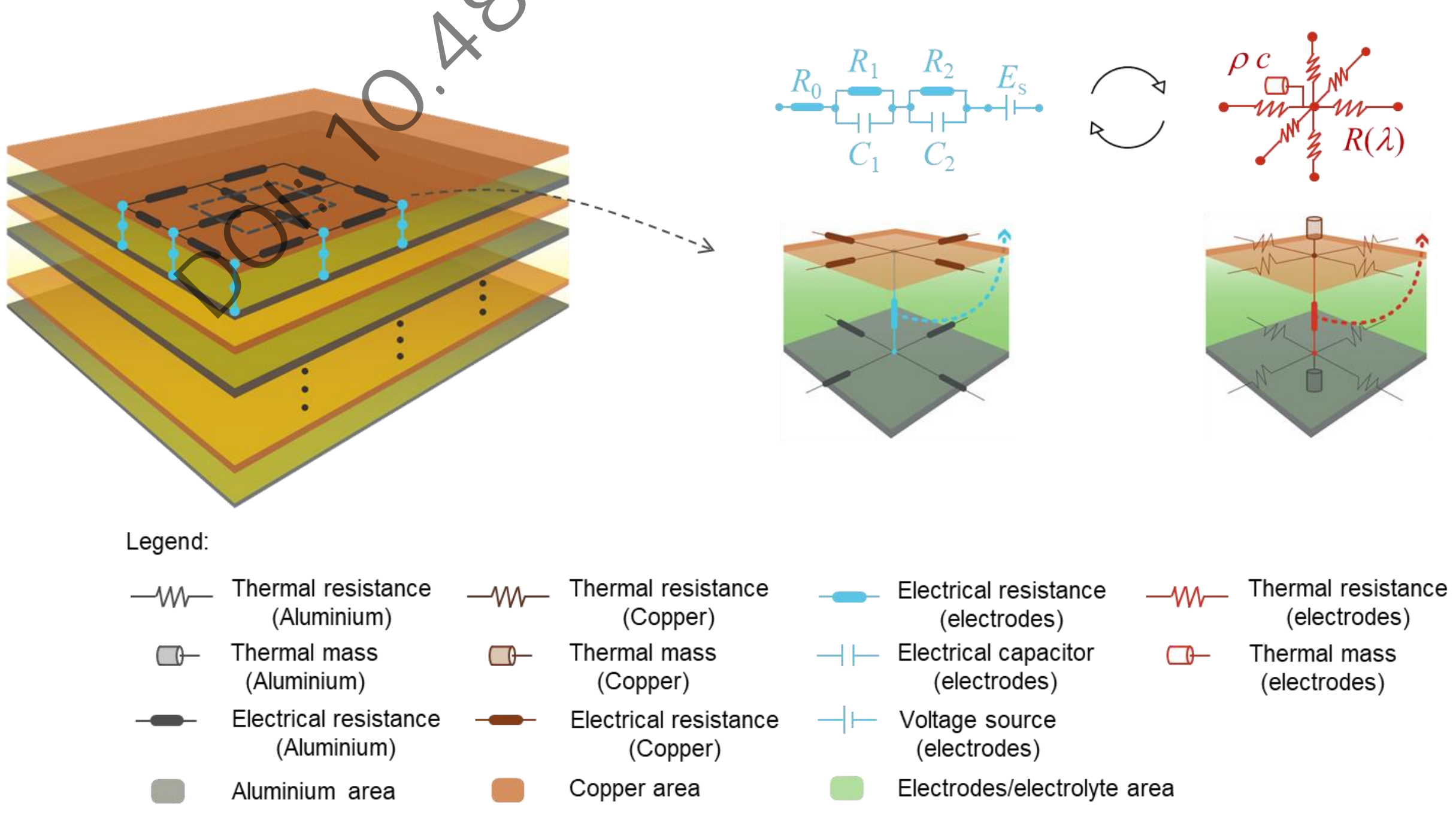

Fig. 5 Schematic representation of the electro-thermal coupled 3D-ECN for a pouch cell. The grey and orange

regions represent the computational domain of aluminium and copper current collectors, respectively. The blue and red networks constitute the electrical and thermal models for each electrode pair subdomain.

The 3D ECN modelling requires parameters for electrical and thermal units as well as geometrical information. As demonstrated previously, the ECM parameters of electrical unit come from Section 2.2. The geometries of the Kokam 7.5Ah Pouch cell were measured in [35], which is summarised in Table 1 (See Appendix B). Thermal properties of the layer component were also reported in [35] and the density of each component was measured in [36]. These values are summarised and presented in Appendix B ( Table 2 ).

# 3 Results and discussion

This section presents the simulations from the computational framework for constant current discharge scenarios at different operating conditions, and compares them with experimental data. For comparative study, we also conducted simulations with a DFN model that is coupled with a classical lumped thermal model. As discussed in the introduction, accurately solving a fully coupled thermal-electrochemical 3D model is a formidable challenge. Therefore, it is common to apply lumped DFN-thermal model to simulate the performance of a commercialised cell [37–39]. The comparative study here was therefore conducted to examine the applicability of this commonly-used approach and also to demonstrate the advantages of computational framework over it.

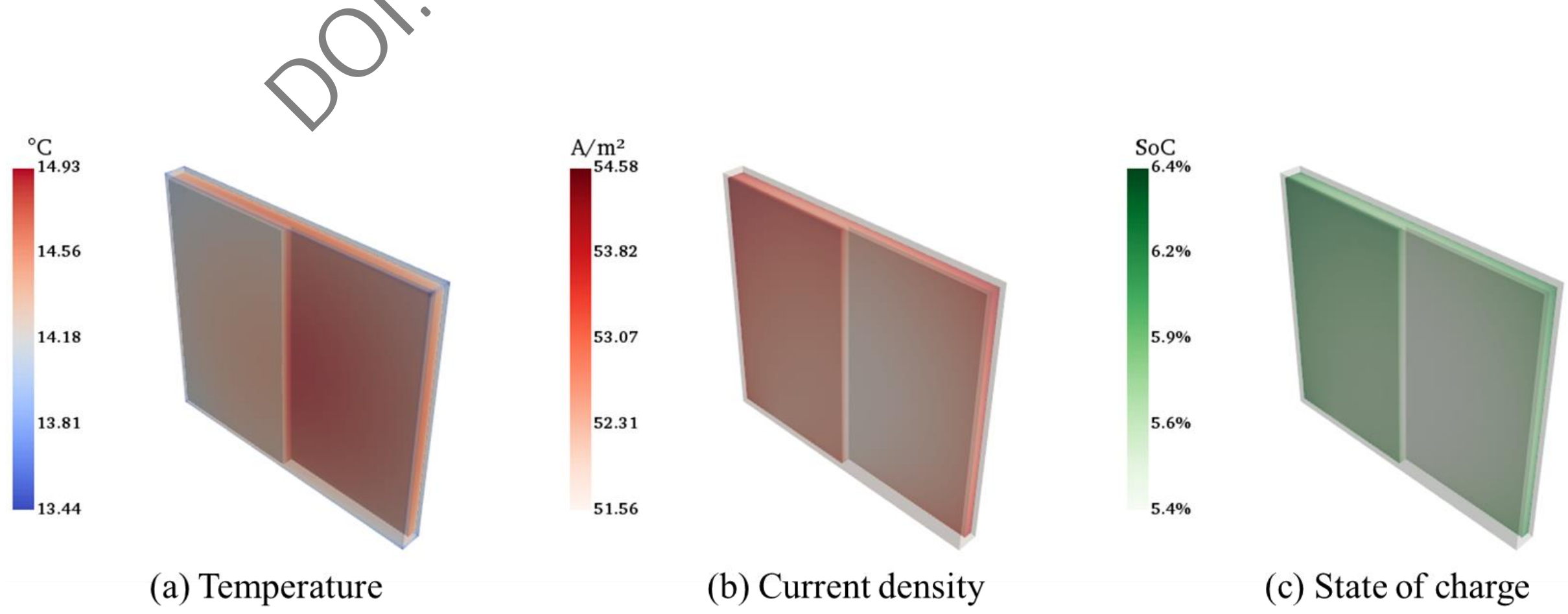

(a) Temperature (b) Current density (c) State of charge

Fig. 6 A snapshot(At t = 3590 s) of the computational framework simulate at the condition of 1C and $T_{am}$ = 55°C. (a) temperature, (b) current density, (c) SoC

As an example, Fig. 6 presents an snapshot of simulation through our computational framework for a Kokam 7.5 Ah cell that discharges at the condition of 3C and $T_{am}$ = 5°C. The cooling scheme is forced convective cooling, same as described in the experiment of [9] and the convective heat transfer coefficient on the surface of the cell was set as 35 W/(m$^2$K) ($h_{conv}$). The internal heat generation leads to pronounced temperature increase, consequently reaching to a value much higher than the ambient 5°C. It is seen that there is significant inhomogeneity across the layer due to surface cooling [40,41]. The temperature difference across the cell can reach to 1.6 °C. Similar non-uniformity can be observed in electrochemical quantities. The largest current density within the cell can reach to 54.58 A/m$^2$, whereas the smallest is 51.56 A/m$^2$. For SoC, the difference between the largest and smallest is 1%. This inhomogeneity influences the overall performance of Li-ion batteries significantly and has been discussed extensively [13,40].

### 3.1 Effect of C-rate

Fig. 7(a) compares the simulations of the computational framework and the lumped DFN-thermal model for constant current discharge at two C-rates (1C and 3C with $T_{am}$=25 °C), with the experimental data [9] added for reference. It is seen that at 1C the difference between the framework and the DFN model is negligible. Both models have a satisfactory agreement with experimental data and the predicting error is no more than 37 mV. However, at 3C, the computational framework better predicts experimental data than the DFN model when discharging capacity is larger than 5 Ah. The largest predicting error of computational framework is 35 mV, whereas that of the lumped DFN model is 93 mV.

The framework outperforms the lumped DFN model at high C-rate scenarios when the internal heat generation is high. The predicted cell average temperature is demonstrated in Fig. 7(b). At 1C, the heat generation is relatively low ( Fig. 7(c) ). The heat generated internally can dissipate more easily and cell temperature would be close to the ambient ($T_{am}$ = 25 °C), where the thermal effects are very small. The difference of cell average temperature predicted by the framework and DFN model is approximately 1 °C, which doesn't lead to much difference in voltage prediction. At 3C, due to the $I^2$ law, the heat generate is much higher ( Fig. 7(c) ), and the thermal effects become very significant, leading to large inhomogeneity across the cell. The distributed ECN model well calculates this inhomogeneity, the temperature difference within the cell reaches up to 1.7 °C (The largest and lower temperature in the cell is depicted by the upper and lower edge of $\Delta T$ shown in Fig. 7(b)). By comparison, the DFN-thermal model is not able to capture this inhomogeneous feature, due to its lumped feature. Accounting for the non-uniformity can lead to a better prediction of the cell overall performance [13]. At the end of discharge, the cell average temperature predicted by the framework is 31.5 °C, 4.2 °C higher than that predicted by the lumped DFN model. Since the overall internal resistance of cell decreases with temperature ( as shown in Fig. 4), a higher prediction of cell temperature can therefore lead to a smaller prediction of voltage drop, resulting in a better agreement with experimental data.

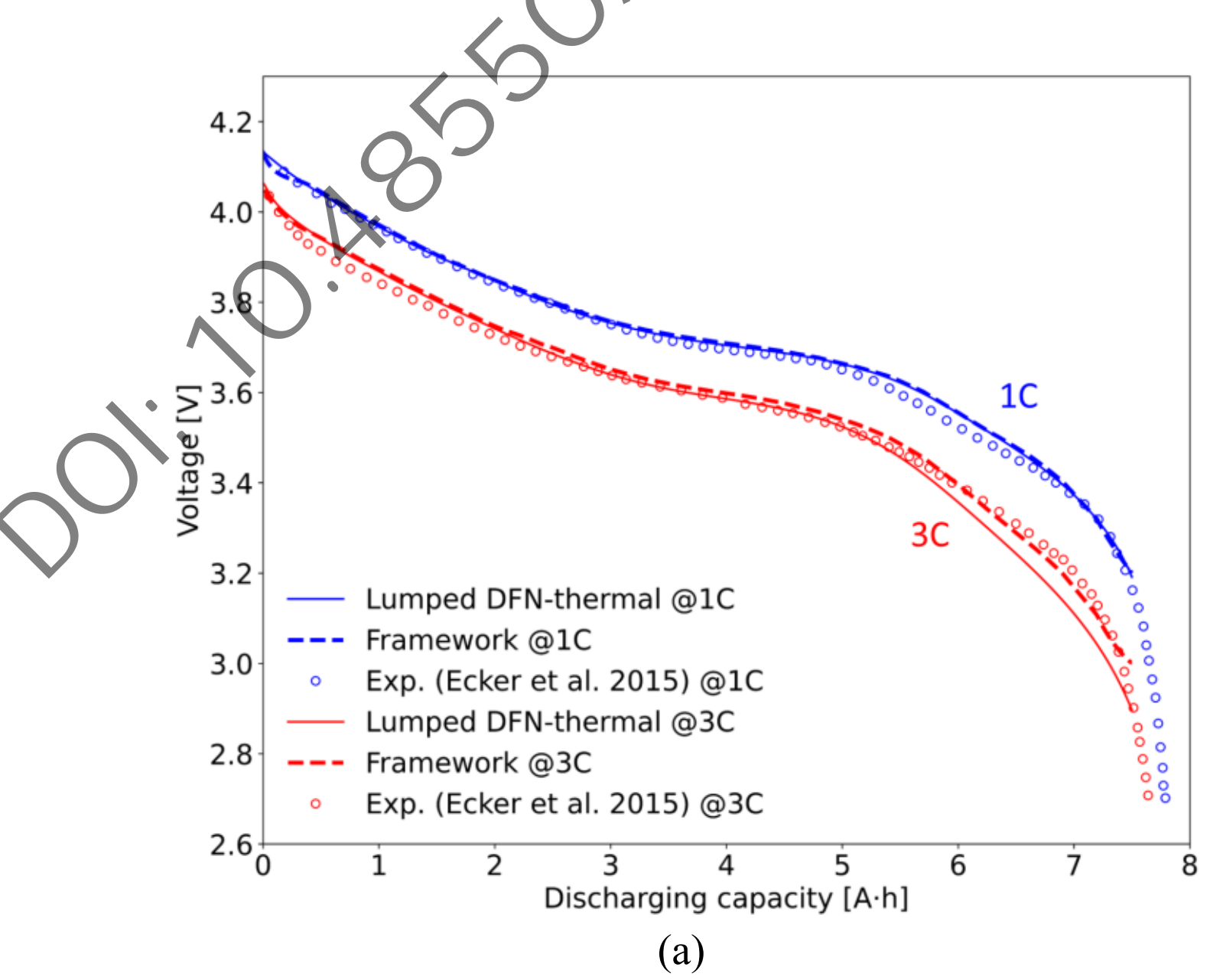

(a)

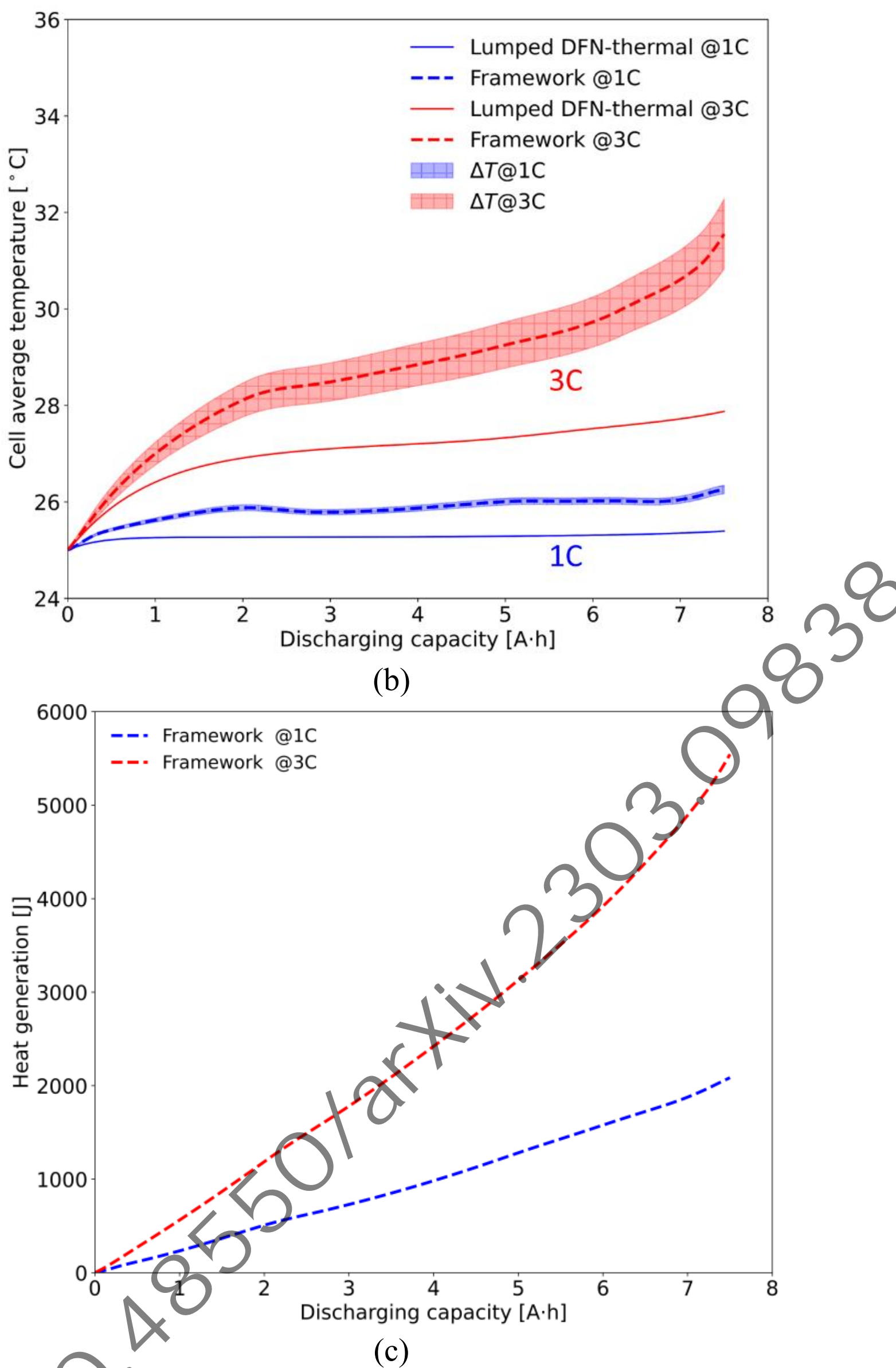

Fig. 7 Comparison of the computational framework and the lumped DFN model for constant current discharge at $T_{am}$=25 °C; Results at 1C and 3C are shown in blue and red respectively. (a) voltage vs. discharging capacity (experimental data are added for reference); (b) cell average temperature vs. discharging capacity(the upper and lower edges of $\Delta T$ depict the highest and lowest temperature in the cell calculated by the computational framework); (c) Heat generation vs. discharging capacity calculated by the computational framework

## 3.2 Effect of temperature

Fig. 8(a) compares the simulations of the computational framework and the lumped DFN model for 1C constant current discharge at three different temperatures, with experimental data [9] added for reference. At $T_{am}$= 25 and 40 °C, the predictions of the framework and the DFN model are very close, the difference between them is less than 10 mV. Both of them show reasonable agreement with the experimental data. However, at $T_{am}$= 0 °C, the DFN model starts to under-predict voltage from1600 s, whereas the framework demonstrates satisfactory predicting capacity until 3200 s. Before 3200 s, the largest predicting error of the framework is 33 mV while that of DFN is 97 mV. The off-prediction of the framework after 3200 s could come from the parametrisation of electrode diffusivity used in the DFN model [9] that the framework is based on. The electrode diffusivity in the DFN model was parametrised based on the assumption that the electrode has a same particle size [9]. However, the measurements in the literature [30] shows that the particle sizes of both electrodes in the Kokam 7.5 Ah cell have a bimodal distribution rather than a unimodal distribution [42]. The application of single-particle size model in this case can lead to off-prediction, especially at the end of discharge at high C-rates [10].

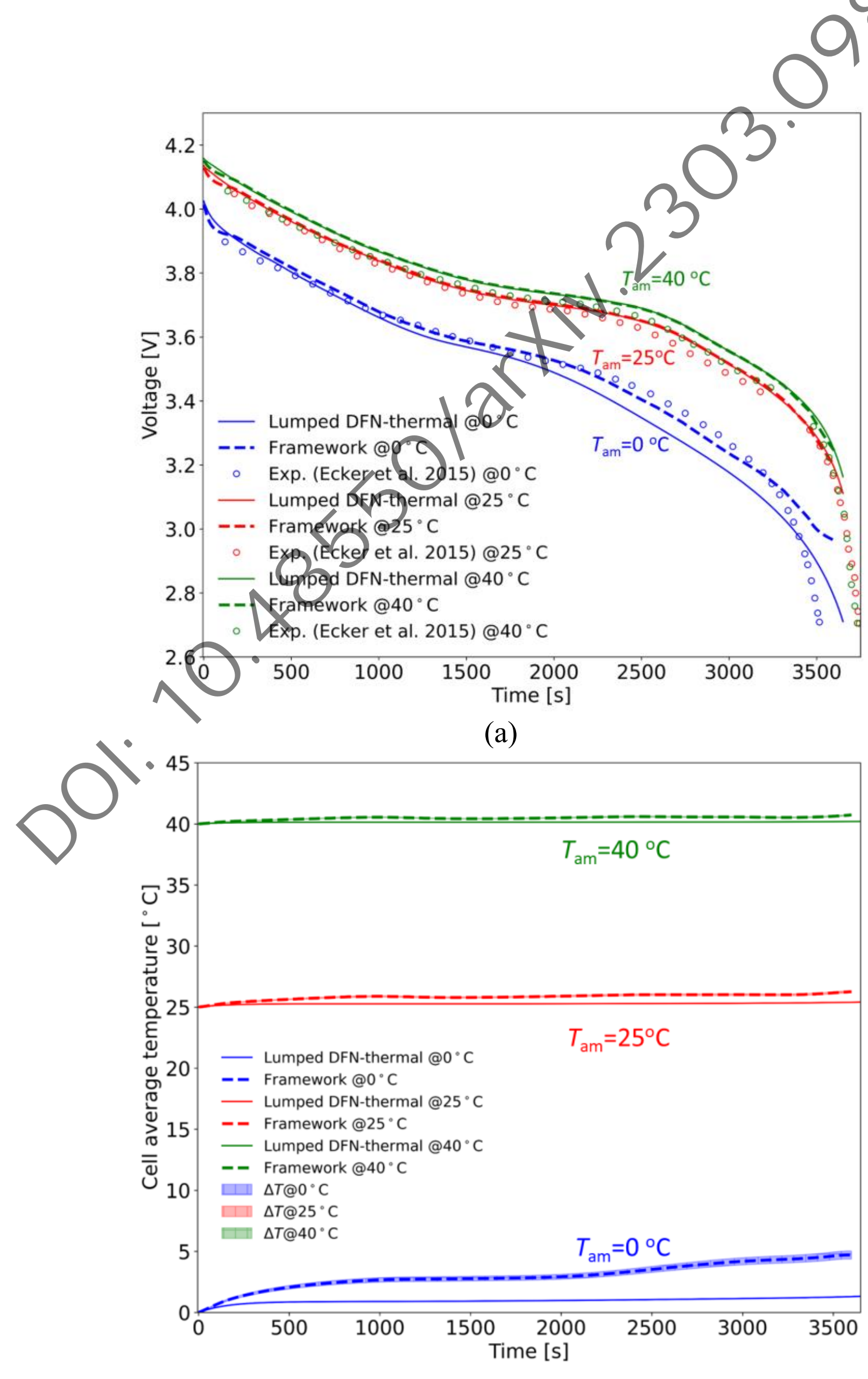

(a)

(b)

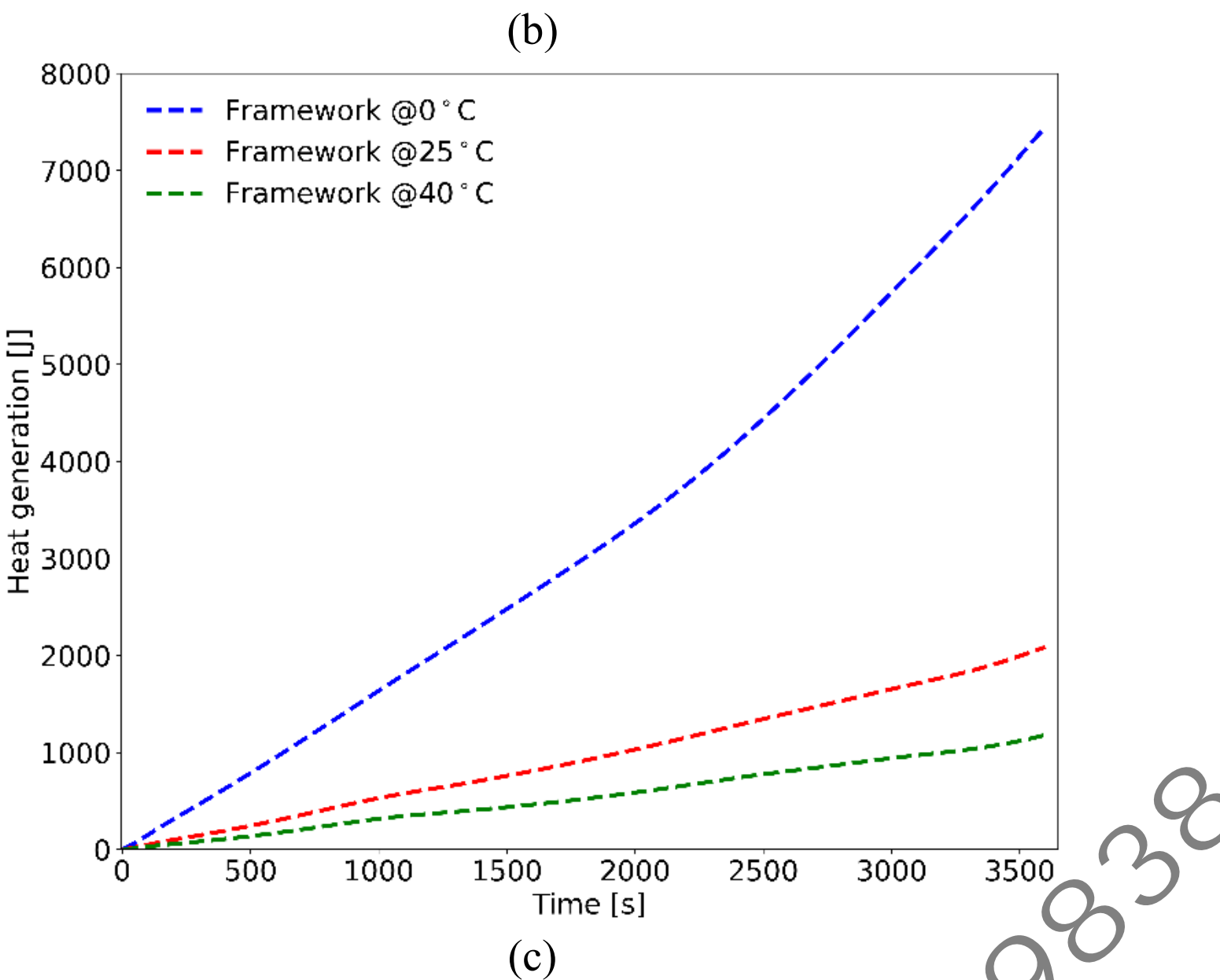

(c)

Fig. 8 Comparison of the computational framework and the lumped DFN model for constant current discharge at 1C-rate ; Results at 0°C, 25°C, and 40 °C are shown in blue, red, and green respectively; (a) voltage vs. time (experimental data are added for reference [9]); (b) cell average temperature vs. time(the upper and lower edges of $\Delta T$ depict the highest and lowest temperature in the cell calculated by the computational framework); (c) Heat generation vs. time calculated by the computational framework

To conclude, at low-temperature ambient, the framework is betters than DFN model. This is also because of a better cell temperature prediction as shown in Fig. 8(b). At $T_{am}$= 25 and 40 °C, the overall resistance of the cell is relatively small and heat generation are not significant ( Fig. 8(c) ). For the same reason as discussed in Section 3.2, the thermal effects are negligible and the predictions of the DFN model don't deviate much from the framework ( Fig. 8(b) ). However, at low-temperature ambient, resistance increases drastically (see Fig. 4 in Section 2.3) , leading a much larger heat generation. With a more accurate consideration of these thermal effects, the framework has a better prediction of cell average temperature than the DFN model. At the end of discharge, the framework has a prediction of 4.7 °C, 3.4 °C higher than that predicted by DFN model. Due to steep gradient at low-temperature range (See Fig. 4), the overall cell resistance (dominated by $R_0$ and $R_2$) is very sensitive to temperature. Therefore, a relatively small improvement in temperature prediction can lead to obvious difference in voltage prediction.

### 3.3 Further discussions

When the cell is operating with both high C-rate and low temperature, thermal effects become vitally important. Fig. 9 compares computational framework and the lumped DFN model for constant current discharge at 3C (Lack of experimental data for reference). According to the calculations of the framework, At $T_{am}$= 5 ºC, the temperature difference within the cell can reach up to 2.2 ºC. With the consideration of the inhomogeneity across the cell, the cell average temperature predicted by the framework is significantly higher than that predicted by the DFN model ( Fig. 9(b) ). For the same reason as explained in Section 3.1 and 3.2, the voltage prediction by the framework is therefore pronouncedly higher than that of the lumped DFN model. The predicting difference can be up to 377 mV.

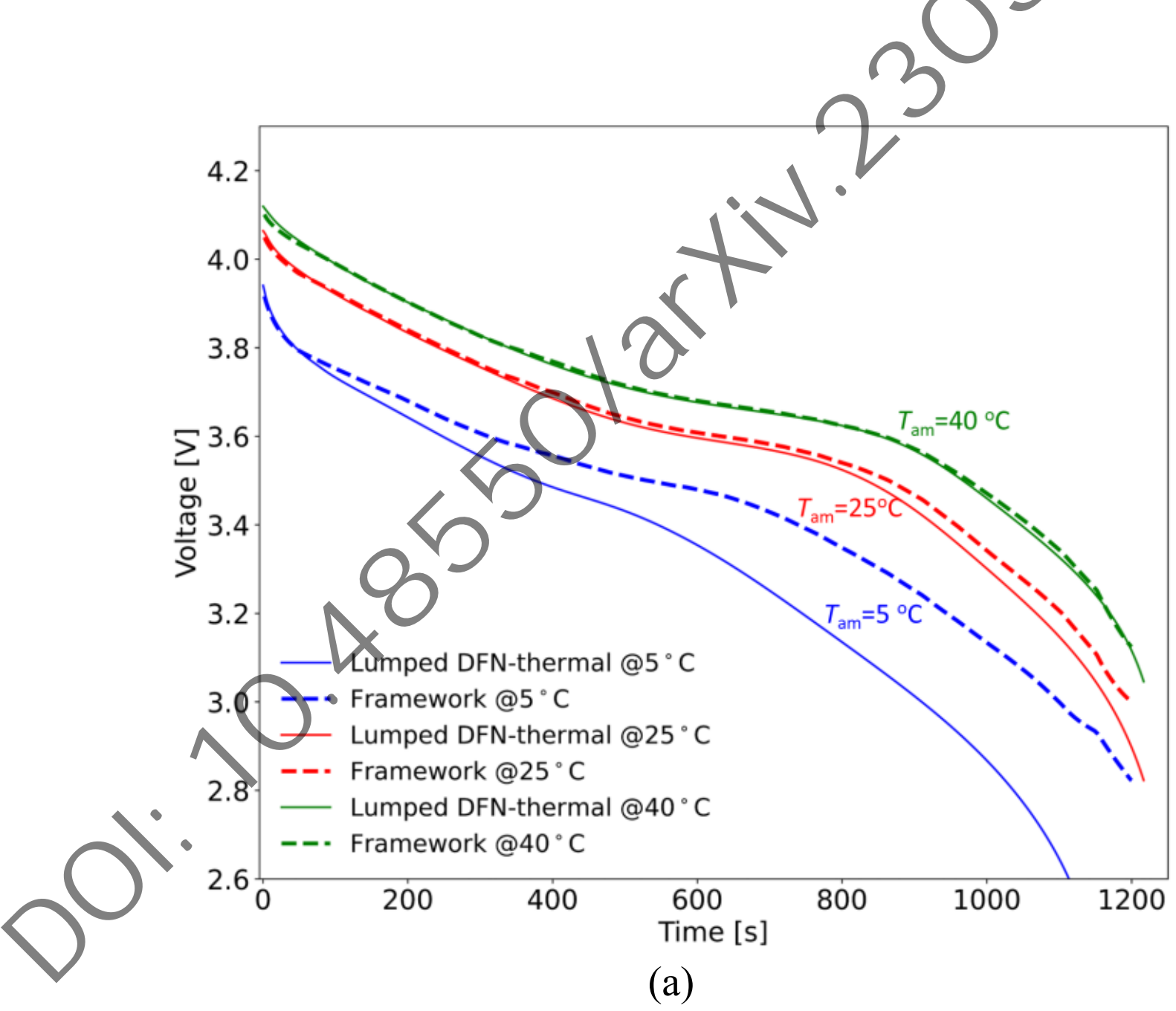

(a)

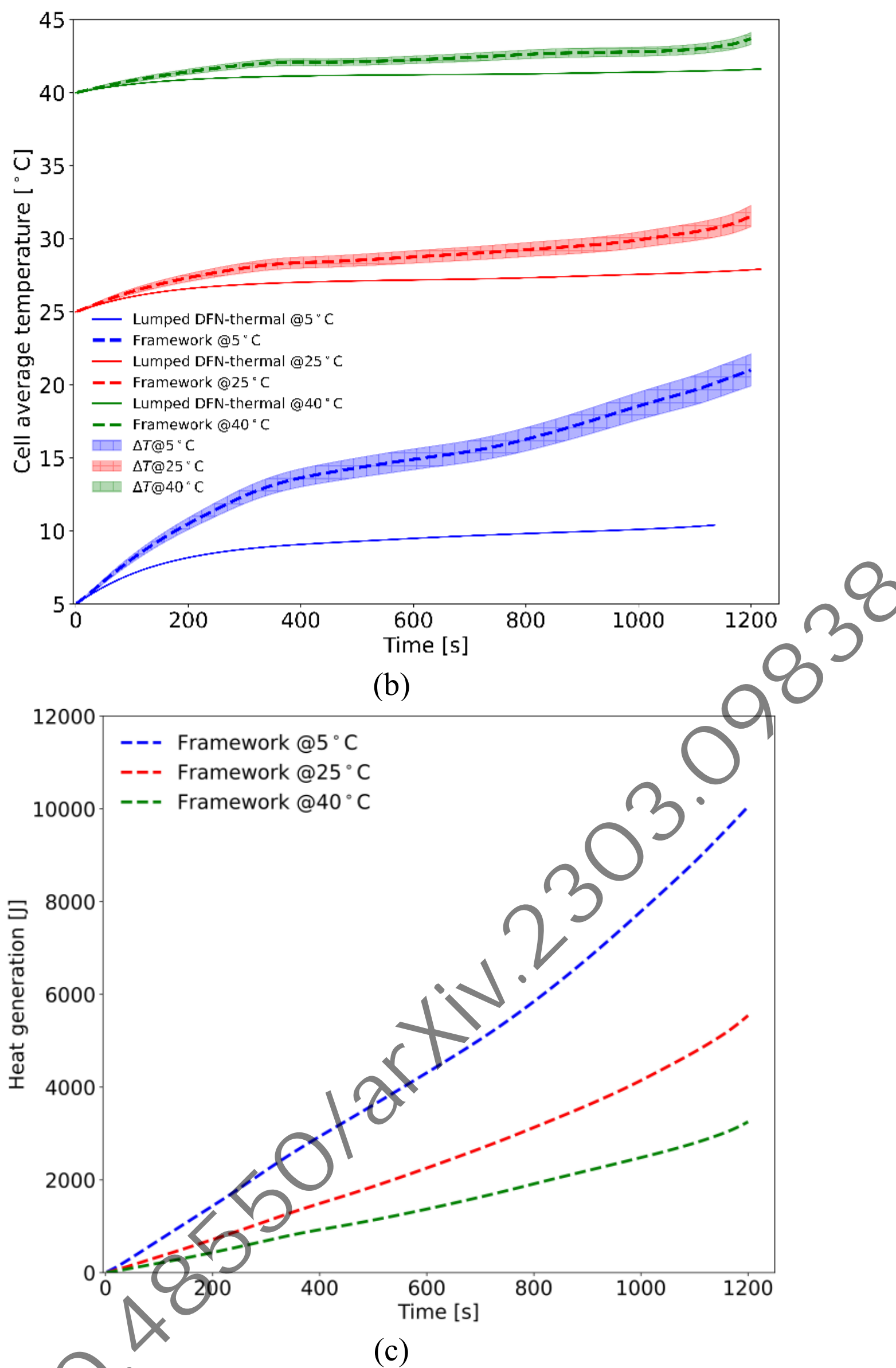

Fig. 9 Comparison of the computational framework and the lumped DFN model for constant current discharge at 3 C-rate; Results at 5°C, 25°C, and 40 °C are shown in blue, red, and green respectively; (a) voltage vs. time; (b) cell average temperature vs. time(the upper and lower edges of $\Delta T$ depict the highest and lowest temperature in the cell calculated by the computational framework); (c) Heat generation vs. time calculated by the computational framework

It is seen that, DFN model coupled with a classical lumped thermal model is not able to accurately predict the cell-level performance at high C-rate or low-temperature scenarios. This is because that the applicability of the classical lumped thermal model depends on several assumptions. However, some assumptions do not apply here.

First, the lump model is limited to the problem with Biot number less than 0.1 [43]. We calculated the Biot number of the Kokam 7.5Ah cell in thickness direction, which is the direction that has the smallest Biot number. The calculation was through Eq.(1).

$$\mathrm{Bi} = \left(\sum_i \frac{\delta_\mathrm{i}}{k_i} N_i\right) h_{\mathrm{conv}} \tag{1}$$

$h_{\mathrm{conv}}$ is the convective heat transfer coefficient on the surface of the cell, the value is 35 W/(m$^2$K) (See the first paragraph of Section 3) . δ is the thickness of the component, *k* is the heat conductivity of it, and *N* is the number of layers of the component. The subscript *i* refers to different components in the cell, including current collectors, separator, anode, cathode, and casing. All of these parameters has been summarised in Table 1. The Biot number in thickness direction is 0.45, much higher than the 0.1 criterion. Secondly, the classical lumped model deviates from reality when there is internal heat generation. The larger heat generation is, the less accurate the simulation would be [44]. When the cell is operating at high C-rate and or low-temperature scenarios, the large amount of internal heat generation can result in inaccurate modelling results.

Currently, large-format batteries are gaining popularity in electric vehicles and energy storage systems [21,45]. Compared to the Kokam 7.5Ah pouch cell investigated in this study, large-format batteries often exhibit greater thermal gradients and higher internal temperatures during cycling [46,47]. The lumped DFN-thermal model becomes even less suitable for predicting the performance of these cells, underscoring the necessity for the proposed computational framework.

# 4 Conclusions

In this research, we build a computational framework that integrates DFN model and 3D distributed ECN model (PyECN) to simulate the performance of Li-ion batteries. DFN model is employed as an ideal experimental rig to avoid the thermal inhomogeneity across the cell in the realistic experiments and produce characterisation data in a much more swiftly. The generated characterisation data are then fitted to obtain ECM parameters at different C-rates and temperatures. Finally, the performance simulations are conducted using PyECN, where the ECM parameters are used as input.

The validity of the computational model is examined by simulating constant current discharge experiments [9] of a Kokam 7.5Ah pouch cell at different operating conditions and by comparing it with the commonly-used lumped DFN-thermal coupled model. It is found that at low-C rates and high temperatures, the performance of the computational framework and the lumped DFN model is similar. However, when it comes to low-temperature or high C-rate scenarios, the computational framework outperforms the lumped DFN model significantly. The largest prediction error of the computational model is 37 mV at 3 C-rate &$T_{am}$ = 25$^{\circ}$C, and 33 mV (before 3200s) at 1 C-rate &$T_{am}$ = 0 $^{\circ}$C. By comparison, the largest prediction error of the lumped DFN model is 93 mV at 3 C-rate &$T_{am}$ = 25$^{\circ}$C, and 97 mV (before 3200 s) at 1 C-rate &$T_{am}$ = 0 $^{\circ}$C. This is because thermal effects become significant at low-temperature or high C-rate scenarios due to large generation, profoundly affecting the cell performance. The computational framework can reasonably calculate this electrochemical-electro-thermally coupling effect, whereas for the lumped DFN model, the underpinning assumptions no longer apply in those scen. When the cell is discharged at both high C-rate and low-temperature ambient, the thermal effects become even more pronounced, leading to a considerable difference between the framework and the DFN model. At 3 C-rate &$T_{am}$ = 25$^{\circ}$C, the voltage predicted by the two models is up to 377 mV.

As the energy and power densities of Li-ion batteries continue to rise, accurately predicting their performance becomes increasingly challenging due to significant internal thermal effects. By Combining the strengths of DFN model and distributed ECN model, our computational framework can simulate the complicated interplay between electrochemistry, thermal process, and electricity within a cell, while keeping the computational resources and possessing time at a low level. We expect that this computational approach will stand out as a efficient toolset, assisting researchers and engineers in the design and control of Li-ion battery cells and packs.

# Appendix A

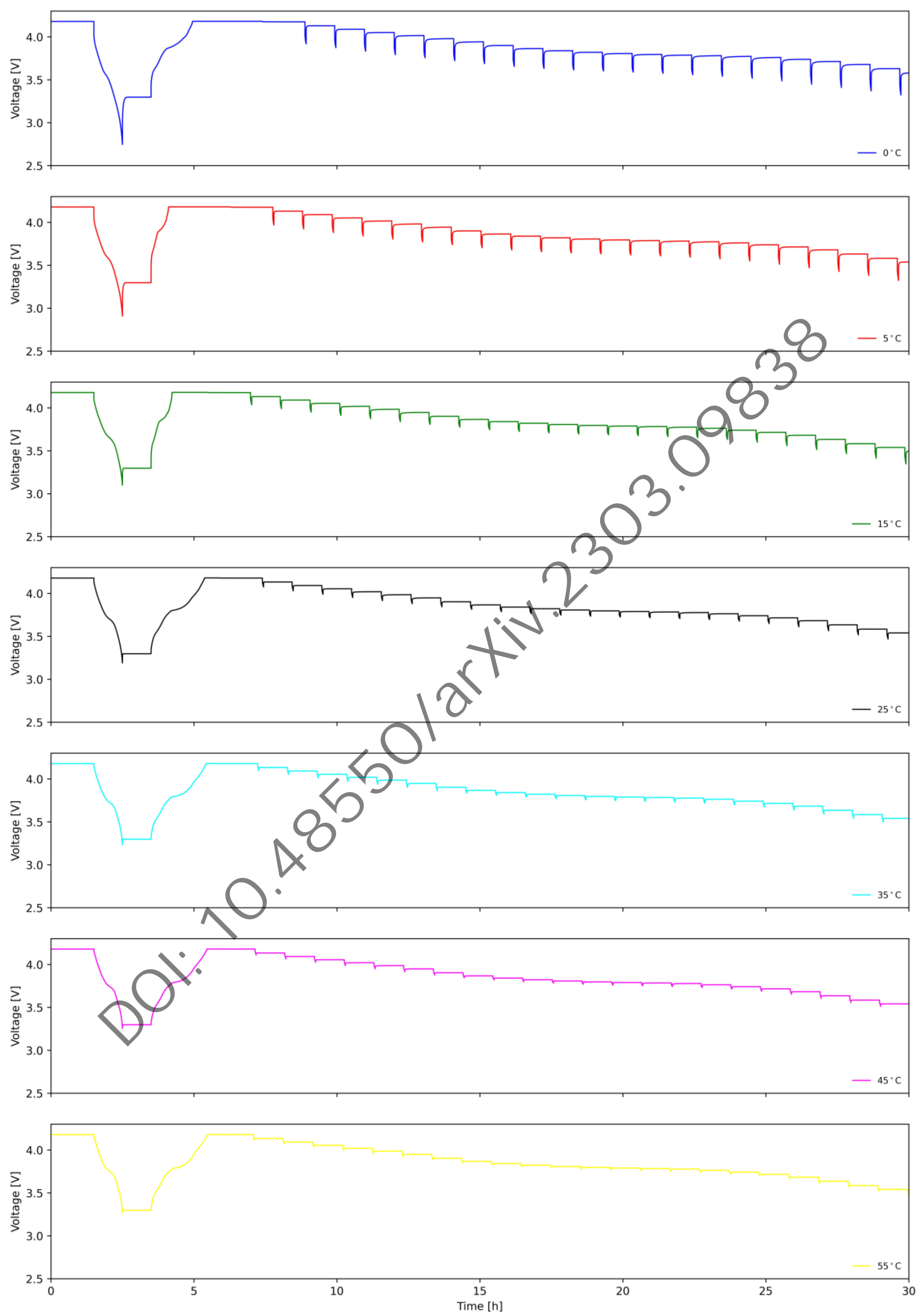

Fig. 10 1C discharge characterization data generated from PyBaMM at different operating temperatures ($T_{amb}$ = 0 °C, 5 °C, 15 °C, 25 °C, 35 °C, 45 °C, 55 °C); A set of data consists of constant current data and GITT data.

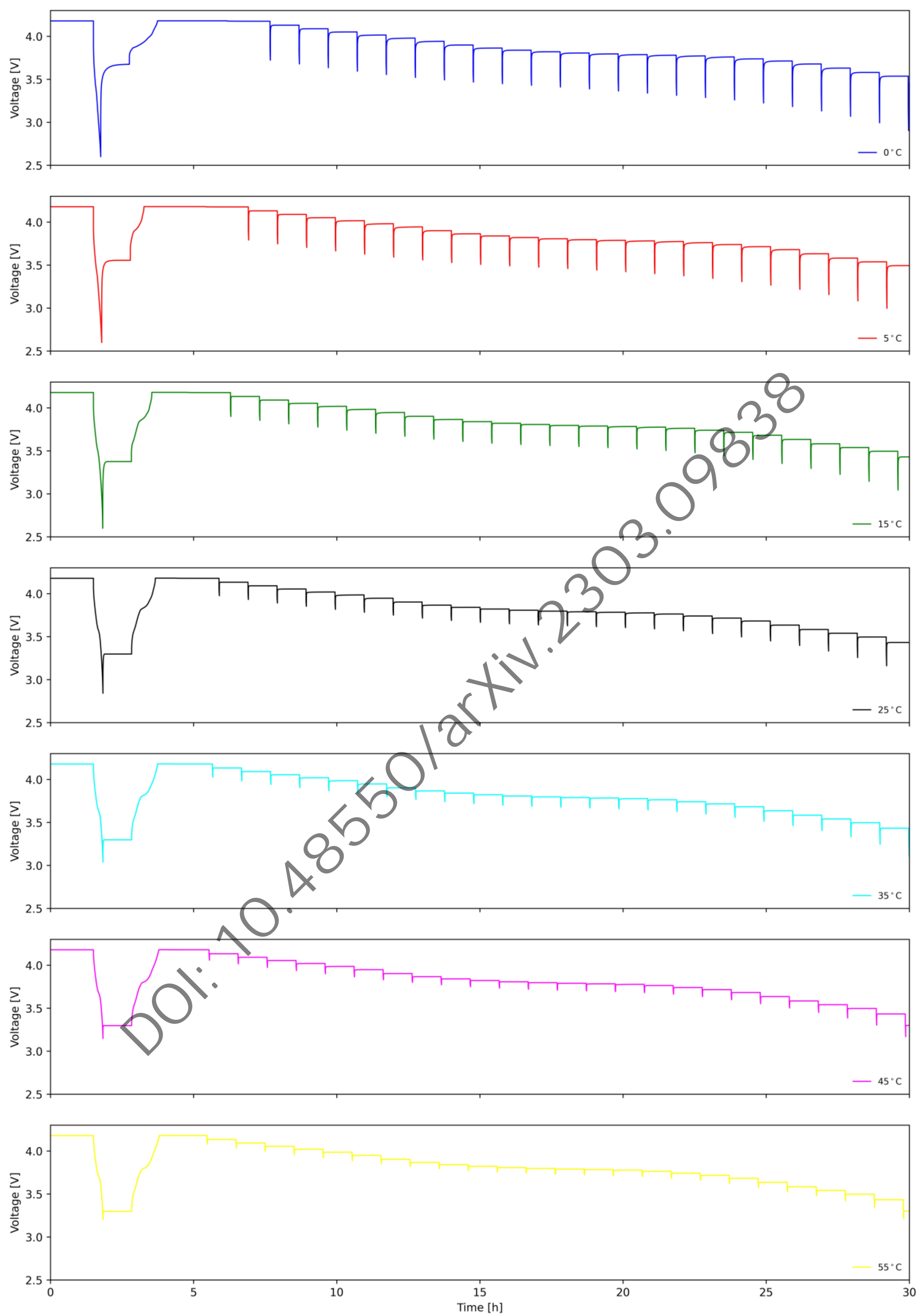

Fig. 11 3C discharge characterization data generated from PyBaMM at different operating temperatures ($T_{amb}$ = 0 °C, 5 °C, 15 °C, 25 °C, 35 °C, 45 °C, 55 °C); A set of data consists of constant current data and GITT data.

# Appendix B

Table 1 Dimension of Kokam 7.5 Ah cell measured in [35]

| Parameter | Value |
|---|---|
| Cell length (mm) | 89.5 |
| Cell width (mm) | 101.5 |
| Cell thickness (mm) | 7.4 |
| Negative tab width (mm) | 7.0 |
| Negative tab thickness (mm) | 0.2 |
| Positive tab width (mm) | 6.9 |
| Positive tab thickness (mm) | 0.2 |
| Negative tab internal length (mm) | 10 |
| Positive tab internal length (mm) | 10 |
| Tab locations (on the cell) | Same end |
| Negative Tab Position (width dimension) | 4.5 mm offset |
| Positive Tab Position (width dimension) | 30.9 mm offset |
| Negative Tab Position (Thickness) | Fully offset |
| Positive Tab Position (Thickness) | Fully offset |

Table 2 The layer component and thermal properties for Kokam 7.5Ah; *CC represents Current collector [35,48].

| Component | Negative CC* | Positive CC | Separator | Anode | Cathode | Casing |
|---|---|---|---|---|---|---|
| Heat conductivity ($W.m^{-1}K^{-1}$) | 398 | 238 | 0.33 | 1.045 | 0.44 | 238 |
| Density ($kg/m^3$) | 8940 | 2700 | 1063 | 1909 | 4000 | 2700 |
| Thickness per layer (mm) | 0.0147 | 0.0151 | 0.0190 | 0.0737 | 0.0545 | 0.1600 |
| Number of layers | 24 | 25 | 54 | 50 | 50 | 2 |
| Volumetric proportion of cell | 4.53% | 4.66% | 11.72% | 45.46% | 33.62% | 3.77% |

# Credit author statement

**Han Yuan**: Conceptualization, Modelling, Validation, Formal analysis, Investigation, Writing, and Review. **Shen Li**: Modelling, Modelling discussion, writing, and Review. **Tao Zhu**: Modelling discussion, and Review. **Simon O'Kane**: Modelling discussion and Review. **Carlos Garcia**: Discussion and Review. **Gregory J. Offer**: Conceptualization, Funding acquisition, Supervision, and Review. **Monica Marinescu**: Conceptualization, Funding acquisition, Supervision, and Review.

## Declaration of competing interest

The authors declare that they have no known competing financial interests or personal relationships that could have appeared to influence the work reported in this paper.

## Acknowledgments

The research has received funding from the Innovate UK through the GENESIS project (grant number 10007488) and EPSRC Faraday Institution Multi-Scale Modelling project (EP/S003053/1, grant number FIRG003). The authors also acknowledge the contribution of Amir Amiri in creating the image in Figure 5.